\documentclass[aps,prb, twocolumn, superscriptaddress,floatfix,longbibliography]{revtex4-2}

\newcommand{\dd}{\mathrm{d}}
\newcommand{\tr}{\mathrm{tr}}
\newcommand{\sgn}{\mathrm{sgn}}
\newcommand{\order}[1]{\mathcal{O}\!\left(#1\right)}

\usepackage{amsmath,amssymb,amsfonts}
\usepackage{graphicx}
\usepackage{slashed}
\usepackage{bm}
\usepackage{xcolor}
\usepackage{geometry}
\usepackage{hyperref}
\usepackage{listings}
\usepackage{color}
\usepackage{subcaption}
\usepackage{caption}
\usepackage{booktabs}
\usepackage{comment}

\hypersetup{
    colorlinks=true,
    urlcolor=blue,
    citecolor=blue,
    linkcolor=blue
}

\begin{document}

\title{From Dirac Cones to Semions: An Exact Finite-Size Theory of Parity-Anomaly Transport in Chiral Spin Liquids}

\author{Kumar Ghosh}
\email{jb.ghosh@outlook.com}
\affiliation{E.ON Digital Technology, Laatzener Str.\ 1, 30539 Hannover, Germany}

\begin{abstract}
Chiral spin liquids realize a topological state whose universal
response is a fractional spin Hall conductance $\nu_s$.  The three
quantities that determine this response, the integer Chern number of
the fractionalized spinons, the level of the emergent Chern--Simons
gauge field, and the physically measured spin pump, are related but
distinct, and their relation is often stated only schematically.  Here
we derive it from a single object: the parity-odd determinant of a
gapped Dirac cone on a spatial cylinder, resummed exactly to all
orders in the compact holonomy.  This determinant fixes the map from
spinon topology to measurable response, and proves that finite-size
corrections to the topological pump are strictly exponential, with no
universal $1/L$ term.  We test the resulting predictions on the
kagome chiral spin liquid at three independent levels: the exact
one-loop field theory, a parton band-structure calculation
($C=-1$, converging exponentially over cylinders four to twelve sites
wide), and an interacting density-matrix renormalization group flux
pump on the explicitly chiral $J$--$J_\chi$ Hamiltonian
($\nu_s=-0.500\pm0.011$).  All three agree with the analytic prediction
without adjustable parameters, providing a fully quantitative bridge
between microscopic topology and observable fractional response.
\end{abstract}

\maketitle

%=======================================================================
\section{Introduction}
\label{sec:introduction}
%=======================================================================

A chiral spin liquid (CSL) is a gapped quantum magnet with intrinsic
topological order and spontaneously broken time-reversal symmetry, but
without conventional magnetic order~\cite{KalmeyerLaughlin1987,
WenWilczekZee1989}.  The Kalmeyer--Laughlin state is the bosonic analogue
of a Laughlin state at filling $\nu=1/2$; its universal properties
include semionic bulk excitations, two ground states on a torus, a
single chiral boson at the edge, and a half-quantized response to a
conserved spin $U(1)$ probe.  These signatures have been identified
numerically in kagome models using exact diagonalization, DMRG,
entanglement spectra, modular matrices, and adiabatic flux insertion
\cite{Gong2014,He2014,Bauer2014,Hu2015,HeBhattacharjee2016}.

The continuum origin of the parity-odd response is the $(2+1)$-
dimensional parity anomaly of massive two-component spinons
\cite{NiemiSemenoff1983,Redlich1984PRL,Redlich1984PRD}.  A single
continuum cone determines only a half-integer infrared contribution
and the quantized jump across a mass inversion.  Large-gauge
invariance requires a lattice ultraviolet completion, whose integer
part is fixed by the Wilson regulator or, equivalently, by the
remote filled bands~\cite{CosteLuscher1989}.  On a compact direction, one
must additionally distinguish a local derivative coefficient from the
complete holonomy-dependent determinant.  The latter, rather than a
finite-order derivative expansion, has the correct behavior under
large gauge transformations \cite{DeserGriguoloSeminara1997,
DeserGriguoloSeminara2002,AitchisonFosco1997,FoscoRossiniSchaposnik1998}.

Four quantities enter the response of a CSL and are frequently
conflated in the literature: the integer Chern number $C$ of the
occupied spinon band, the level $K_{\rm em}$ of the emergent
Chern--Simons gauge field, the physical spin Hall invariant $\nu_s$
that a flux pump measures, and the chiral central charge $c_-$
of the edge theory.  These are related by the parton projection and by
the topological field theory, but they are not the same integer.
The organizing goal of this paper is to derive their relation as a
theorem rather than assert it as a dictionary, and to verify the
resulting predictions at three independent levels of description.

\emph{Analytic result.} We compute the full mass-dependent
parity-odd two-point kernel of a Dirac cone on
$\mathbb R_\tau\times\mathbb R_x\times S^1_L$, keeping every external
compact harmonic.  For a closed transverse two-manifold with quantized
flux we integrate the differential response exactly in the compact
holonomy, obtaining the mass-dependent determinant for this background
class.  The Redlich half-level and the Coste--Luescher integer class
restore large-gauge invariance in an explicit way.  The
fixed-holonomy correction is a sum of virtual winding
contributions proportional to $e^{-rL/\xi}$, so the finite-size
correction is exponential and there is no universal $1/L$ term at
fixed nonzero gap.  Combined with the Green-function bridge to the
Bloch Chern number and the hydrodynamic $K$-matrix reduction, this
yields the universal dictionary
$K_{\rm em}=2C$, $\nu_s=C/2$, $c_-=\sgn C$ for the semion CSL.

\emph{Numerical validation on the kagome lattice.} We construct the
Abrikosov-fermion mean-field Hamiltonian on the kagome lattice with
uniform complex hopping, compute the Chern number, and evaluate the
finite-cylinder response for YC-$L_y$ cylinders with $L_y=4$--$12$.
The lattice response converges exponentially to $C_{\rm occ}=-1$ with
no $1/L$ correction, confirming the central analytic prediction.  A
pilot interacting DMRG calculation on the explicitly chiral kagome
model with $J_\chi/J=0.25$ then measures $\nu_s=-0.500\pm0.011$ over
two flux periods, with the correct $4\pi$ semionic periodicity and a
uniform bulk scalar chirality throughout the scan.  The three
determinations (field theory, parton lattice, interacting DMRG) form a
quantitatively closed loop that fixes the topological content of the
kagome CSL as the $U(1)_{-2}$ Chern--Simons theory with $\nu_s=-1/2$.

The remainder of the paper is organized as follows.
Section~\ref{sec:framework} introduces the four bookkeeping quantities
and the normalization conventions.  Section~\ref{sec:direct} derives
the exact cylinder determinant.  Section~\ref{sec:parton} converts
spinon Chern bands to the semion topological field theory using the
Green-function bridge and the hydrodynamic $K$-matrix.
Section~\ref{sec:manybody} defines the interacting many-body pump.
Sections~\ref{sec:kagome_realization} and~\ref{sec:dmrg} carry out the
noninteracting and interacting kagome validations.
Section~\ref{sec:discussion} synthesizes the three levels and
discusses open directions.  Appendices collect technical
derivations, benchmark models, and comparisons with the Kitaev phase
and with distinct finite-size scaling laws.

%=======================================================================
\section{Framework: four quantities and their relations}
\label{sec:framework}
%=======================================================================

We use Euclidean coordinates $x^\mu=(\tau,x,y)$ with
$y\equiv y+L$ and $\epsilon^{\tau x y}=+1$.  For a compact $U(1)$
field $b$ with unit minimal charge,
\begin{equation}
 S_{\rm CS}[b]=\frac{i k}{4\pi}\int b\wedge\dd b
 =\frac{i k}{4\pi}\int\dd^3x\,
 \epsilon^{\mu\nu\rho}b_\mu\partial_\nu b_\rho.
 \label{eq:CS_convention}
\end{equation}
For a cone of probe charge $q_v$ coupled to $a$, we introduce
$b_v\equiv q_v a$ so that
\begin{equation}
 \Theta_v=\oint_{S^1}b_v,\qquad
 \mathcal N_v=\frac{1}{2\pi}\int_{\Sigma_2}\dd b_v\in\mathbb Z,
 \label{eq:normalized_global_data}
\end{equation}
where $\Sigma_2$ is a closed Euclidean two-manifold.  All
large-gauge statements below are made in terms of $b_v$; the factor
$q_v^2$ is restored at the end.

The four bookkeeping quantities are
\begin{align}
 & C_\alpha :\ \text{integer Chern number of spinon species }\alpha,
 \label{eq:def_C}\\
 & K_{\rm em} :\ \text{level of the dynamical emergent gauge field},
 \label{eq:def_K}\\
 & \nu_s =t^T K^{-1}t:\ \text{physical spin Hall invariant},
 \label{eq:def_nu}\\
 & c_- :\ \text{chiral central charge controlling }\kappa_{xy}/T.
 \label{eq:def_c}
\end{align}
In our Euclidean convention, integrating out the internal gauge fields
produces $S_{\rm eff}[A^s]=-i\nu_s(4\pi)^{-1}\int A^s\dd A^s$; the sign
of a reported pump depends on the orientation of the seam and cut, and
we choose orientations so that positive $C$ gives $\Delta S^z=+C/2$.

For the semion CSL obtained from two spin species with
$C_\uparrow=C_\downarrow=C=\pm1$, the derivation in
Sec.~\ref{sec:parton} will yield
\begin{equation}
 \begin{aligned}
 K_{\rm em}&=2C, & \nu_s&=\frac{1}{2C}=\frac{C}{2},\\
 |\det K|&=2, & c_-&=\sgn C.
 \end{aligned}
 \label{eq:dictionary_semion}
\end{equation}
An emergent $U(1)_2$ theory therefore has physical spin response of
magnitude $1/2$, not $2$, and one chiral boson with $|c_-|=1$.  These
identifications are the target of the rest of the paper: they are
consequences of the parity anomaly and the parton constraint, not
independent conventions.

%=======================================================================
\section{Exact parity-odd response on the cylinder}
\label{sec:direct}
%=======================================================================

\subsection{Massive cone and background holonomy}
\label{subsec:action}

Consider one two-component Dirac cone $v$ with Euclidean action
\begin{equation}
 \begin{aligned}
 S_v={}&\int_0^L\dd y\int\dd\tau\,\dd x\,
 \bar\psi_v\bigl[
 \gamma^\tau D_\tau+v_{\perp,v}\gamma^x D_x\\
 &\hspace{15mm}+v_{\parallel,v}\gamma^yD_y+m_v\bigr]\psi_v,
 \end{aligned}
 \label{eq:Dirac_action}
\end{equation}
with $D_\mu=\partial_\mu+i q_v a_\mu$.  The compact boundary condition
is
\begin{equation}
 \psi_v(y+L)=e^{i\alpha_v}\psi_v(y),
 \qquad \alpha_v\in[0,2\pi),
 \label{eq:spin_structure}
\end{equation}
so that the total phase seen by the cone is
\begin{equation}
 \vartheta_v=\alpha_v+\Theta_v,
 \qquad \Theta_v=q_v\oint_{S^1}a_y\dd y=\oint_{S^1}b_v,
 \label{eq:total_holonomy}
\end{equation}
and the internal compact momentum is
\begin{equation}
 k_{y,n}=\frac{2\pi n+\vartheta_v}{L},\qquad n\in\mathbb Z.
 \label{eq:twist}
\end{equation}
On a microscopic cylinder, $\alpha_v$ may include the valley momentum
projected onto the wrapping vector as well as the chosen spin
structure.  We define the effective circumference, correlation
length, and dimensionless product
\begin{equation}
 L_v^{\rm eff}=\frac{L}{|v_{\parallel,v}|},\quad
 \xi_v=\frac{|v_{\parallel,v}|}{|m_v|},\quad
 \lambda_v=\frac{L}{\xi_v}=|m_v|L_v^{\rm eff}.
 \label{eq:xi}
\end{equation}
The sign $\chi_v=\pm1$ combines the orientation of the linearized
Bloch map with the chosen irreducible gamma-matrix representation, fixed
operationally so that the occupied cone contributes
$C_v=\chi_v\sgn(m_v)/2$ before ultraviolet completion; the mapping is
made explicit in Appendix~\ref{app:DiracChern}.

\subsection{Full mass-dependent kernel}
\label{subsec:full_kernel}

The quadratic effective action is
\begin{equation}
 \Gamma_v^{(2)}[a]=\frac12\sum_p
 a_\mu(-p)\Pi_v^{\mu\nu}(p)a_\nu(p),
 \label{eq:Gamma2}
\end{equation}
with rescaled external momentum
\begin{equation}
 \widetilde p^2=p_\tau^2+v_{\perp,v}^2p_x^2+v_{\parallel,v}^2p_y^2,
 \quad p_y=\frac{2\pi r}{L},\quad r\in\mathbb Z.
 \label{eq:p_tilde}
\end{equation}
At quadratic order the loop-momentum terms cancel identically in the
parity-odd numerator.  Consequently, even though compactification
breaks Euclidean rotational symmetry, the mass-dependent odd tensor
is purely antisymmetric and has the exactly transverse form
\begin{equation}
 \Pi_{v,{\rm odd}}^{\mu\nu,\rm IR}(p)=
 \frac{i q_v^2}{2\pi}\,
 \mathcal K_v^{\rm IR}(p;L,\vartheta_v)\,
 \epsilon^{\mu\nu\rho}p_\rho .
 \label{eq:Pi_full}
\end{equation}
The central result of this section (derived in
Appendix~\ref{app:full_kernel}) is that for every Euclidean
noncompact momentum and compact harmonic $r$,
\begin{equation}
 \boxed{\begin{aligned}
 \mathcal K_v^{\rm IR}(p;L,\vartheta_v)
 &=\frac{\chi_vm_v}{2}\int_0^1\frac{\dd u}{\Delta_{v,u}}\\
 &\times
 \frac{\sinh(L_v^{\rm eff}\Delta_{v,u})}
 {\cosh(L_v^{\rm eff}\Delta_{v,u})-
  \cos(\vartheta_v+2\pi r u)}
 \end{aligned}}
 \label{eq:K_full}
\end{equation}
with
\begin{equation}
 \Delta_{v,u}=\sqrt{m_v^2+u(1-u)\widetilde p^2}.
 \label{eq:Delta_u}
\end{equation}
No derivative expansion has been made.  The Ward identity holds
algebraically:
\begin{equation}
 p_\mu\Pi_{v,{\rm odd}}^{\mu\nu,\rm IR}(p)=0.
 \label{eq:Ward_exact}
\end{equation}
A gauge-invariant ultraviolet regulator adds only a momentum-local
constant to the odd form factor,
\begin{equation}
 \Pi_{v,{\rm odd}}^{\mu\nu,\rm reg}
 =\frac{i q_v^2}{2\pi}
 \left[\mathcal K_v^{\rm IR}+n_v+\frac{\chi_v\eta_v}{2}\right]
 \epsilon^{\mu\nu\rho}p_\rho,
 \label{eq:Pi_regulated}
\end{equation}
where $\eta_v=\pm1$ labels the parity-anomaly branch and
$n_v\in\mathbb Z$ labels an integer regulator class.  In a multicone
lattice theory the cone-wise decomposition of the ultraviolet term is
bookkeeping dependent; only the total regulated level and the
mass-inversion jumps are invariant.
Equation~\eqref{eq:K_full} is complete in momentum dependence for the
mass-dependent one-loop contribution; Eq.~\eqref{eq:Pi_regulated}
displays one convenient cone-wise representation of the local
ultraviolet completion.

Two limits illuminate the structure.  In the decompactification
limit,
\begin{equation}
 \mathcal K_v^{\rm IR,\infty}(p)=
 \frac{\chi_vm_v}{2}\int_0^1
 \frac{\dd u}{\sqrt{m_v^2+u(1-u)\widetilde p^2}},
 \label{eq:K_decompactified}
\end{equation}
the standard nonlocal massive-Dirac form factor, tending to
$\chi_v\sgn(m_v)/2$ at zero momentum.  The finite-$L$ residual has
the exact winding expansion
\begin{align}
 \mathcal K_v^{\rm IR}-\mathcal K_v^{\rm IR,\infty}
 ={}&\chi_vm_v\sum_{\ell=1}^{\infty}\int_0^1
 \frac{\dd u}{\Delta_{v,u}}
 e^{-\ell L_v^{\rm eff}\Delta_{v,u}}\notag\\
 &\times\cos\!\left[\ell(\vartheta_v+2\pi r u)\right],
 \label{eq:K_winding_expansion}
\end{align}
so exponential suppression at fixed nonzero mass is manifest before
the zero-momentum limit is taken.  Restricting further to $r=0$ and
$p_\tau,p_x\to 0$,
\begin{align}
 \mathcal K_v^{\rm IR}(0;L,\vartheta_v)
 &=\frac{\chi_v}{2}\sgn(m_v)
 R(\lambda_v,\vartheta_v),
 \label{eq:K_local_IR}\\
 R(\lambda,\vartheta)
 &\equiv\frac{\sinh\lambda}
 {\cosh\lambda-\cos\vartheta},
 \label{eq:Rdef}
\end{align}
so that, in terms of the original field $a$,
\begin{equation}
 \boxed{
 k_{v}^{\rm IR}(L,\vartheta_v)=
 \frac{q_v^2}{2}\chi_v\sgn(m_v)
 R(\lambda_v,\vartheta_v)}.
 \label{eq:k_local_exact}
\end{equation}
Because compactification is spatial and the physical temperature is
zero, the $r=0$ retarded correlator is analytic near
$(\omega,p_x)=(0,0)$ whenever the sub-band gap
\begin{equation}
 \Delta_{L,v}=\min_n\sqrt{m_v^2+
 \left(\frac{2\pi n+\vartheta_v}{L_v^{\rm eff}}\right)^2}>0
 \label{eq:subband_gap}
\end{equation}
is nonvanishing.  After analytic continuation
$p_\tau\to-i(\omega+i0^+)$, no spectral weight occurs below the
two-particle threshold $2\Delta_{L,v}$: the static and transport
limits commute for this gapped zero-temperature cylinder.  This
statement fails at a gap closing and does not extend to finite
physical temperature, where thermal occupation can generate
Landau-damping nonanalyticities.  The gauge-invariant lattice analysis
of Karthik and Narayanan likewise separates a local regulator term
from a nonlocal finite-mass continuum
contribution~\cite{KarthikNarayanan2015}.

\subsection{Exact holonomy resummation}
\label{subsec:exact_holonomy}

The nonlinear determinant is not known for an arbitrary gauge
background, but is exactly calculable for a constant compact holonomy
and quantized field strength on a closed transverse two-manifold
$\Sigma_2$.  Using the normalized field $b_v=q_v a$ of
Eq.~\eqref{eq:normalized_global_data},
\begin{equation}
 \Theta_v=\oint_{S^1}b_v,\qquad
 \mathcal N_v=\frac{1}{2\pi}\int_{\Sigma_2}\dd b_v\in\mathbb Z.
 \label{eq:transverse_flux}
\end{equation}
This is the spatially compact counterpart of the exact backgrounds
studied at finite temperature~\cite{AitchisonFosco1997,
FoscoRossiniSchaposnik1998}.  In the pump interpretation,
$\mathcal N_v$ is an auxiliary transverse twist used to probe the
Berry curvature, distinct from the physical path that varies only the
cylinder-threading flux.

A normalization subtlety must be handled carefully
(Appendix~\ref{app:holonomy}).  Evaluating the Chern--Simons
functional first on a general field and only then restricting to
constant $b_y$ and transverse flux, the two terms related by
integration by parts contribute equally, giving
\begin{equation}
 S_{\rm CS}[b_v]\big|_{\Theta_v,\mathcal N_v}
 =i k\,\mathcal N_v\Theta_v,
 \label{eq:CS_restricted}
\end{equation}
not the factor-of-two-smaller value obtained by substituting a
strictly constant $b_y$ before taking the zero-momentum limit.

Introduce the continuous-branch primitive
\begin{equation}
 \mathcal A_{\alpha}(x,\Theta)
 =\int_0^{\Theta}\frac{\dd\phi}{2}
 \frac{\sinh x}{\cosh x-\cos(\alpha+\phi)},
 \label{eq:A_holonomy_integral}
\end{equation}
whose closed form
\begin{align}
 \mathcal A_{\alpha}(x,\Theta)
 &=\operatorname{Cont}\!\Bigg\{
 \arctan\!\left[\coth\frac{x}{2}
 \tan\frac{\alpha+\Theta}{2}\right]\notag\\
 &\hspace{15mm}-\arctan\!\left[\coth\frac{x}{2}
 \tan\frac{\alpha}{2}\right]\Bigg\}
 \label{eq:A_holonomy_closed}
\end{align}
follows the branch continuously as $\Theta$ varies.  The mass-odd
infrared determinant is
\begin{equation}
 \boxed{
 \Gamma_{v,{\rm odd}}^{\rm IR}
 =i\chi_v\mathcal N_v\,
 \mathcal A_{\alpha_v}(m_vL_v^{\rm eff},\Theta_v)}.
 \label{eq:Gamma_exact_IR}
\end{equation}
Its derivative reproduces the differential infrared level via
\begin{equation}
 \frac{\partial\mathcal A_{\alpha}(x,\Theta)}{\partial\Theta}
 =\frac12\frac{\sinh x}{\cosh x-\cos(\alpha+\Theta)}.
 \label{eq:A_derivative}
\end{equation}
The specialization to antiperiodic ($\alpha=\pi$) and periodic
($\alpha=0$) spin structures, and the resulting connection to the
familiar $\tanh(|m|L/2)$ coefficient~\cite{BabuDasPanigrahi1987}, is
given in Appendix~\ref{app:holonomy}.

The infrared determinant winds under large gauge transformations,
\begin{equation}
 \mathcal A_{\alpha}(x,\Theta+2\pi)
 -\mathcal A_{\alpha}(x,\Theta)=\pi\sgn(x).
 \label{eq:A_winding}
\end{equation}
For an odd number of cones this mass-dependent piece is therefore not
the complete gauge-invariant answer.  A gauge-invariant regularization
adds the mass-independent parity-anomaly branch $\eta_v=\pm1$ and an
integer universality-class shift $n_v\in\mathbb Z$; when several cones
are present these labels need not be assigned uniquely cone by cone, and
only the total is physical:
\begin{equation}
 \boxed{\begin{aligned}
 \Gamma_{v,{\rm odd}}^{\rm full}
 &=i\chi_v\mathcal N_v
 \left[\mathcal A_{\alpha_v}(m_vL_v^{\rm eff},\Theta_v)
 +\frac{\eta_v}{2}\Theta_v\right]\\
 &\quad+i n_v\mathcal N_v\Theta_v.
 \end{aligned}}
 \label{eq:Gamma_exact_full}
\end{equation}
Under $\Theta_v\to\Theta_v+2\pi$,
\begin{equation}
 \Delta\Gamma_{v,{\rm odd}}^{\rm full}
 =i\pi\chi_v\mathcal N_v[\sgn(m_v)+\eta_v]
 +i2\pi n_v\mathcal N_v\in2\pi i\mathbb Z,
 \label{eq:large_gauge_check}
\end{equation}
so $e^{-\Gamma}$ is invariant.  Equation~\eqref{eq:Gamma_exact_full}
is the explicit compact-space realization of the Redlich parity anomaly
combined with the integer Coste--Luescher ambiguity.

\subsection{Differential response, cycle average, and no universal $1/L$}
\label{subsec:slope_pump}

For the normalized field $b_v$, define the differential level in the
flux sector $\mathcal N_v\ne0$ by
\begin{equation}
 k_{v}^{\rm diff}(L,\Theta_v)
 \equiv\frac{1}{i\mathcal N_v}
 \frac{\partial\Gamma_{v,{\rm odd}}^{\rm full}}{\partial\Theta_v}.
 \label{eq:kdiff_definition}
\end{equation}
Equation~\eqref{eq:Gamma_exact_full} gives
\begin{equation}
 k_{v}^{\rm diff}(L,\Theta_v)=
 n_v+\frac{\chi_v\eta_v}{2}
 +\frac{\chi_v}{2}\sgn(m_v)
 R(\lambda_v,\alpha_v+\Theta_v).
 \label{eq:k_full_local}
\end{equation}
The cycle average is
\begin{align}
 \overline k_v
 &\equiv\frac{1}{2\pi}\int_0^{2\pi}\dd\Theta_v\,
 k_v^{\rm diff}(L,\Theta_v)\notag\\
 &=n_v+\frac{\chi_v}{2}[\eta_v+\sgn(m_v)]\in\mathbb Z,
 \label{eq:cycle_average_integer}
\end{align}
using $(2\pi)^{-1}\int_0^{2\pi}\dd\Theta\,R(\lambda,\alpha+\Theta)=1$,
which is the Poisson average identity
\begin{equation}
 \frac{1}{2\pi}\int_0^{2\pi}\dd\Theta\,
 R(\lambda,\alpha+\Theta)=1.
 \label{eq:Poisson_average}
\end{equation}
For an independently regularized block, the integer in
Eq.~\eqref{eq:cycle_average_integer} is its Chern number.  In a
physical multicone lattice model the same statement applies to the
sum over all cones and remote bands, whose decomposition is not
unique.  In either case this is a \emph{spinon} band invariant; the
conversion to the fractional many-body spin pump requires the parton
projection of Sec.~\ref{sec:parton}.

A second distinction operates at finite size.  A many-body pump at
fixed transverse twist is an integral of the Berry curvature along a
single flux path and is not exactly quantized on a finite torus or
cylinder.  The topological integer is obtained after integrating over
both twists, or equivalently after averaging the pump over the
transverse twist.  In a gapped phase the residual twist dependence
vanishes exponentially with linear size, so a fixed-twist cylinder
pump approaches the topological value in the thermodynamic limit.
Section~\ref{sec:manybody} states this relation precisely.

Turn now to the thermodynamic limit.  For fixed $m_v\ne0$ and fixed
holonomy, the Poisson-kernel expansion
\begin{equation}
 R(\lambda,\vartheta)=
 1+2\sum_{\ell=1}^{\infty}e^{-\ell\lambda}
 \cos(\ell\vartheta)
 \label{eq:R_expansion}
\end{equation}
gives
\begin{align}
 k_{\rm IR}^{\rm loc}(L)-k_{\rm IR}^{\infty}
 &=\sum_v q_v^2\chi_v\sgn(m_v)
 \cos\vartheta_v\,e^{-L/\xi_v}\notag\\
 &\quad+\order{\sum_v e^{-2L/\xi_v}}.
 \label{eq:k_asymptotic}
\end{align}
Equation~\eqref{eq:K_winding_expansion} yields the same conclusion at
nonzero Euclidean momentum: the asymptotic expansion of the gapped
bulk cone contains no universal algebraic term,
\begin{equation}
 \boxed{c_1=0.}
 \label{eq:c1zero}
\end{equation}
This is the paper's central analytic conclusion for finite-size
behavior.  The limit becomes nonuniform as $m_v\to0$, at a compact
zero mode, or when the circumference is not large compared with
$\xi_v$; no exponential-scaling claim is made in those regimes.  A
robust $1/L$ term in a demonstrably gapped calculation must come from
physics absent from the local massive-cone theory, such as an edge
mode, a critical crossover, or a numerical finite-entanglement effect.
The $1/L$ approach to the continuum found by Karthik and Narayanan is
instead a regulator convergence effect whose $\tau^{-3}$ coefficient
reconstructs a local Chern--Simons term, not a universal circumference
correction of a gapped spatial cylinder~\cite{KarthikNarayanan2015}.

%=======================================================================
\section{From spinon bands to semion topological order}
\label{sec:parton}
%=======================================================================

Section~\ref{sec:direct} showed that each Dirac cone contributes a
half-integer $\chi_v\sgn(m_v)/2$ to the parity-odd level.  A physical
band insulator must have an integer response, and this reconciliation
is the content of Redlich's parity anomaly~\cite{Redlich1984PRL,
Redlich1984PRD} and of Coste and Luescher's Wilson-fermion
completion~\cite{CosteLuscher1989}.  For one unit-charge species,
\begin{equation}
 C_\alpha=n_\alpha+\frac12\sum_{v\in\alpha}
 \chi_v\sgn(m_v)\in\mathbb Z,
 \label{eq:lattice_completed_level}
\end{equation}
where the integer $n_\alpha\in\mathbb Z$ is not an infrared fitting
parameter; it is fixed once the complete microscopic band structure
and filling are specified.  For a probe under which species $\alpha$
has charge $q_\alpha$, the quadratic response level is
$k_{\rm probe}=\sum_\alpha q_\alpha^2 C_\alpha$.

\subsection{Green-function bridge to the Bloch Chern number}
\label{subsec:green_bridge}

The Coste--Luescher integer can be recognized directly as a Bloch
invariant through the interacting Green-function
formula~\cite{IshikawaMatsuyama1987}
\begin{equation}
 \begin{aligned}
 N_3[G]=\frac{1}{24\pi^2}\int\dd\omega\,\dd^2k\,
 \epsilon^{\mu\nu\rho}\,
 \tr\!\bigl[&G\partial_\mu G^{-1}G\partial_\nu G^{-1}\\
 &\times G\partial_\rho G^{-1}\bigr],
 \end{aligned}
 \label{eq:Green_invariant}
\end{equation}
where $\mu,\nu,\rho\in\{\omega,k_x,k_y\}$.  For a noninteracting
insulator $G^{-1}=i\omega-h(\bm k)$, closing the frequency contour
gives (Appendix~\ref{app:GreenChern})
\begin{equation}
 C_\alpha=\frac{1}{2\pi}\sum_{n\in{\rm occ}}
 \int_{\rm BZ}\dd^2k\,\Omega_{n\alpha}(\bm k),
 \label{eq:lattice_Chern}
\end{equation}
with
\begin{equation}
 \Omega_{n\alpha}=i\left(
 \langle\partial_{k_x}u_{n\alpha}|\partial_{k_y}u_{n\alpha}\rangle
 -\langle\partial_{k_y}u_{n\alpha}|\partial_{k_x}u_{n\alpha}\rangle
 \right).
 \label{eq:Berry_curvature}
\end{equation}
The Coste--Luescher regulator integer and the remote-band integer are
therefore two representations of the same three-dimensional
Green-function topology.  Equation~\eqref{eq:lattice_Chern} is the
most direct invariant in a microscopic calculation and is the one used
in Sec.~\ref{sec:kagome_realization}.

Near isolated Dirac points, Eq.~\eqref{eq:lattice_completed_level}
becomes
\begin{equation}
 C_\alpha=n_\alpha+\frac12\sum_{v\in\alpha}
 \chi_v\sgn(m_v),
 \label{eq:C_signed_cones}
\end{equation}
which allows cancellation between valleys.  For two valleys $K$ and
$K'$ with $\chi_K=+1$ and $\chi_{K'}=-1$,
\begin{align}
 m_{K'}&=-m_K\equiv-m_H,
 & C_{\rm cones}&=\sgn(m_H),
 \label{eq:Haldane_mass}\\
 m_{K'}&=m_K\equiv m_S,
 & C_{\rm cones}&=0.
 \label{eq:trivial_mass}
\end{align}
The first is a Haldane-type chiral pattern, the second is trivial.  A
full Brillouin-zone calculation remains mandatory because additional
bands or gap closings can change $n_\alpha$.  The Coste--Luescher
Wilson-degree formula is stated in Appendix~\ref{app:GreenChern} for
completeness.

\subsection{Parton projection and the semion \texorpdfstring{$U(1)_{2}$}{U(1)2} theory}
\label{subsec:gauge_projection}

Use the Abrikosov-fermion representation
\begin{equation}
 S_i^a=\frac12 f_{i\alpha}^\dagger\sigma^a_{\alpha\beta}f_{i\beta},
 \qquad \sum_\alpha f_{i\alpha}^\dagger f_{i\alpha}=1,
 \label{eq:parton_rep}
\end{equation}
so that the constraint introduces a compact emergent $U(1)$ gauge
field $a$.  Both spin species have unit emergent gauge charge and spin
charges $s_\uparrow=+1/2$, $s_\downarrow=-1/2$ under the physical probe
$A^s$.

For two filled spinon Chern bands with $C_\uparrow=C_\downarrow=C$, the
topological action in hydrodynamic form is
\begin{align}
 S_{\rm top}={}&\frac{iC}{4\pi}\sum_{\sigma}
 \int\alpha_\sigma\dd\alpha_\sigma
 +\frac{i}{2\pi}\int a\dd(\alpha_\uparrow+\alpha_\downarrow)
 \notag\\
 &+\frac{i}{4\pi}\int A^s\dd(\alpha_\uparrow-\alpha_\downarrow).
 \label{eq:parton_hydrodynamic}
\end{align}
Integrating over $a$ imposes $\dd(\alpha_\uparrow+\alpha_\downarrow)=0$;
setting $\alpha_\uparrow=b$ and $\alpha_\downarrow=-b$ up to a flat
gauge field gives
\begin{equation}
 S_{\rm top}=\frac{i\,2C}{4\pi}\int b\dd b
 +\frac{i}{2\pi}\int A^s\dd b,
 \label{eq:semion_K_action}
\end{equation}
so that
\begin{equation}
 K=(2C),\qquad t=(1),\qquad
 K_{\rm em}=2C.
 \label{eq:Kt}
\end{equation}
The physical spin Hall invariant is
\begin{equation}
 \nu_s=t^TK^{-1}t=\frac{1}{2C}=\frac{C}{2}
 \qquad(C=\pm1).
 \label{eq:Kem_nus}
\end{equation}
Integrating out $b$ produces
$S_{\rm eff}[A^s]=-i\nu_s(4\pi)^{-1}\int A^s\dd A^s$.  The quasiparticle
labelled by $\ell=1$ carries spin charge
$t^TK^{-1}\ell=1/(2C)$ modulo an integer and has exchange phase
$\theta_\ell=\pi\ell^TK^{-1}\ell=\pi/(2C)$: the spin-$1/2$
fractionalization and semionic statistics follow from the same $K$-matrix
data.  Universal quasiparticle data are collected in
Appendix~\ref{app:Kmatrix}.

For the kagome CSL constructed by scalar spin chirality or complex
spinon hopping, the Haldane pattern $m_{K'}=-m_K$ in
Eq.~\eqref{eq:Haldane_mass} gives
\begin{equation}
 C_\uparrow=C_\downarrow=\sgn(m_H),
 \label{eq:kagome_C}
\end{equation}
whenever no compensating ultraviolet integer arises from remote bands.
The Kalmeyer--Laughlin phase corresponds to
$|C|=1$ per spin species.  Reversing scalar chirality reverses $m_H$,
all Chern numbers, the direction of the edge mode, and the sign of the
flux pump.  The same signed-cone test applies to any lattice ansatz
whose low-energy spectrum contains a pair of oppositely oriented Dirac
valleys, but the number of valleys alone does not fix $C$: both the
mass pattern and the full-band Chern number must be verified.  The
universal statement is conditional but sharp:
\begin{equation}
 \begin{gathered}
 C_\uparrow=C_\downarrow=\pm1\\
 \Longrightarrow\quad U(1)_{\pm2}\ \text{semion CSL},
 \quad \nu_s=\pm\frac12.
 \end{gathered}
 \label{eq:universal_semion}
\end{equation}
This derives, rather than assumes, the dictionary quoted in
Sec.~\ref{sec:framework}.  A distinct scenario, the field-induced
non-Abelian Kitaev phase, gives Ising topological order and a chiral
Majorana edge; because generic Kitaev interactions do not conserve
$S^z$, the spin $U(1)$ pump defined below is not generally available,
and the robust response is gravitational (Appendix~\ref{app:kitaev}).

%=======================================================================
\section{Many-body pump and its scaling}
\label{sec:manybody}
%=======================================================================

The parton-level analysis of Sec.~\ref{sec:parton} treats
$C_\uparrow=C_\downarrow$ as noninteracting inputs.  The interacting
counterpart is defined by twisted boundary conditions on the physical
spin Hamiltonian~\cite{NiuThoulessWu1985,Gong2014}.

\subsection{Twisted-boundary many-body Chern number}
\label{subsec:manybody_chern}

For an interacting spin Hamiltonian on a torus, impose twists
$(\theta_x,\theta_y)$ in the conserved $S^z$ channel.  Let
$\{|\Psi_a(\bm\theta)\rangle\}_{a=1}^{d}$ span a topological
ground-state multiplet that remains separated from all excited states
by
\begin{equation}
 \Delta_{\rm mb}(\bm\theta)>0
 \quad\text{for every }(\theta_x,\theta_y)\in T^2.
 \label{eq:twist_gap_condition}
\end{equation}
The non-Abelian Berry connection and curvature are
\begin{align}
 [\mathcal A_i]_{ab}
 &=i\langle\Psi_a|\partial_{\theta_i}\Psi_b\rangle,
 \label{eq:nonabelian_A}\\
 \mathcal F_{xy}
 &=\partial_{\theta_x}\mathcal A_y-
 \partial_{\theta_y}\mathcal A_x-i[\mathcal A_x,\mathcal A_y].
 \label{eq:nonabelian_F}
\end{align}
Only the trace over the entire isolated multiplet is gauge invariant
when states are exchanged under flux insertion.  Its Chern number is
\begin{equation}
 C_{\rm MB}=\frac{1}{2\pi}\int_0^{2\pi}\dd\theta_x
 \int_0^{2\pi}\dd\theta_y\,\tr\mathcal F_{xy}\in\mathbb Z,
 \label{eq:manybody_Chern}
\end{equation}
and for a $d$-fold multiplet sharing the response in the
thermodynamic limit,
\begin{equation}
 \nu_s=\frac{C_{\rm MB}}{d}.
 \label{eq:nu_MB}
\end{equation}
For the semion CSL, $d=2$ and $C_{\rm MB}=\pm1$, so
$\nu_s=\pm1/2$.  A gauge-invariant discrete implementation of
Eq.~\eqref{eq:manybody_Chern} for numerical work is collected in
Appendix~\ref{app:discreteChern}.

At finite size, a pump at fixed transverse twist is not itself the
Chern number.  Define
\begin{equation}
 Q_s(\theta_x)=\frac{1}{2\pi d}\int_0^{2\pi}\dd\theta_y\,
 \tr\mathcal F_{xy}(\theta_x,\theta_y).
 \label{eq:fixed_twist_pump}
\end{equation}
Only its transverse-twist average is exactly topological,
\begin{equation}
 \frac{1}{2\pi}\int_0^{2\pi}\dd\theta_x\,Q_s(\theta_x)
 =\frac{C_{\rm MB}}{d}=\nu_s.
 \label{eq:pump_twist_average}
\end{equation}
For a local gapped Hamiltonian the boundary-twist dependence vanishes
exponentially with linear size, so a fixed-$\theta_x$ pump approaches
$\nu_s$ in the thermodynamic limit.  This is the many-body counterpart
of the distinction between Eqs.~\eqref{eq:k_full_local} and
\eqref{eq:cycle_average_integer}, and it is the observable measured in
Sec.~\ref{sec:dmrg}.

\subsection{Cylinder flux insertion and polarization pump}
\label{subsec:flux_pump}

On an infinite or long cylinder, insert spin flux $\Phi$ through the
hole by twisting all bonds crossing a seam,
$S_i^+S_j^-\mapsto e^{i\Phi}S_i^+S_j^-$.  The spin transferred across
an entanglement cut is either
\begin{equation}
 P_s(\Phi)=\sum_{i\in L}
 \left[\langle S_i^z\rangle_\Phi-\langle S_i^z\rangle_0\right]
 \label{eq:polarization_realspace}
\end{equation}
in real space, or
\begin{equation}
 P_s(\Phi)=\sum_\beta\lambda_\beta^2(\Phi)Q_\beta(\Phi)
 \quad(\mathrm{mod}\ 1)
 \label{eq:polarization_schmidt}
\end{equation}
in terms of the infinite-MPS Schmidt charges.  In the adiabatic
thermodynamic limit, with the seam and cut orientation chosen so that
positive $C$ gives positive pump,
\begin{equation}
 \Delta P_s(2\pi)=\nu_s=\frac{C}{2}.
 \label{eq:half_pump}
\end{equation}
A $2\pi$ cycle maps the identity minimally entangled state to the
semion sector; a second cycle returns to the original sector and
transfers an integer spin.  At finite circumference the fixed-path
value can differ from $C/2$ by exponentially small bulk corrections and
by finite-length, finite-bond-dimension, or nonadiabatic errors, so
both the integrated pump and the local slope
$\partial P_s/\partial(\Phi/2\pi)$ should be reported separately.

\subsection{Prior numerical evidence and target regime}
\label{subsec:priorwork}

Two families of kagome models frame the numerical validation.  Gong,
Zhu, and Sheng studied the time-reversal-invariant $J$--$J'$
Hamiltonian
\begin{equation}
 H=J\sum_{\langle ij\rangle}\bm S_i\cdot\bm S_j
 +J'\sum_{\langle\!\langle ij\rangle\!\rangle}
 \bm S_i\cdot\bm S_j
 +J'\sum_{\langle\!\langle\!\langle ij\rangle\!\rangle\!\rangle}
 \bm S_i\cdot\bm S_j,
 \label{eq:gong_model}
\end{equation}
and found a spontaneously chiral phase over
$0.1\lesssim J'/J\lesssim0.7$; adiabatic flux insertion on
$3\times 24\times 4$ cylinders at $J'/J=0.5$ produced
\begin{equation}
 \Delta S^z_{\rm edge}(2\pi)=\frac12,
 \qquad
 \Delta S^z_{\rm edge}(4\pi)=1,
 \label{eq:gong_pump}
\end{equation}
with the correct semion sector flow~\cite{Gong2014}.  A complementary
explicitly time-reversal-breaking model is
\begin{equation}
 H=J_{\rm HB}\sum_{\langle ij\rangle}\bm S_i\cdot\bm S_j
 +J_\chi\sum_{(ijk)\in\triangle,\nabla}
 \bm S_i\cdot(\bm S_j\times\bm S_k),
 \label{eq:bauer_model}
\end{equation}
with a fixed clockwise orientation of $(i,j,k)$ on every elementary
triangle; Bauer \emph{et al.} identified the CSL phase for
$J_\chi/J_{\rm HB}\gtrsim0.16$ and demonstrated its stability against
Dzyaloshinskii--Moriya coupling, next-nearest-neighbor exchange, and
easy-axis anisotropy~\cite{Bauer2014}.  Because Eq.~\eqref{eq:bauer_model}
explicitly selects one sign of the Dirac mass, it is the cleaner
benchmark for the sign and mass dependence of the local parity-odd
response; the DMRG scan in Sec.~\ref{sec:dmrg} uses this Hamiltonian at
$J_\chi/J=0.25$, well inside the CSL phase.  The strong-coupling
Hubbard origin of Eq.~\eqref{eq:bauer_model} and the historical status
of prior CSL numerics are collected in
Appendix~\ref{app:benchmarks}.

%=======================================================================
\section{Kagome parton band calculation}
\label{sec:kagome_realization}
%=======================================================================

The analytic framework of Secs.~\ref{sec:direct}--\ref{sec:parton}
requires a microscopic lattice model to supply the full-band Chern
number $C$ and the finite-cylinder response.  We provide these from an
Abrikosov-fermion mean-field construction on the kagome lattice with
uniform complex hopping.  Within the projective symmetry group (PSG)
classification of Bieri, Lhuillier, and
Messio~\cite{Bieri2016}, this ansatz belongs to the Kalmeyer--Laughlin
class ($\tau_\sigma=1$, $\tau_R=0$) with nontrivial flux through all
elementary triangles, corresponding to their CSL No.~13 in Table~VI
(PSG No.~3, $\epsilon_2=-1$).  This is the same flux pattern
$[3\phi_1,\pi-6\phi_1]$ found variationally by Hu \emph{et
al.}~\cite{Hu2015} and confirmed by DMRG~\cite{Gong2014,He2014}.

\subsection{Bloch Hamiltonian and gauge convention}
\label{subsec:bloch}

The kagome lattice has three sublattices $A$, $B$, $C$ per unit cell,
with lattice vectors
$\mathbf{a}_1=(1,0)$ and $\mathbf{a}_2=(\tfrac12,\tfrac{\sqrt3}{2})$
in units of the nearest-neighbor distance.  Every nearest-neighbor hop
carries the uniform phase $e^{-i\phi/3}$ in the clockwise sublattice
direction $A\to B\to C\to A$.  Because this phase is the same on every
bond, every elementary triangle (up and down alike) encloses the same
gauge-invariant flux $\phi$.  The resulting $3\times3$ Bloch Hamiltonian
has off-diagonal elements
\begin{align}
 h_{AB}(\mathbf{k})
 &=t\,e^{-i\phi/3}\bigl(1+e^{-i\mathbf{k}\cdot\mathbf{a}_1}\bigr),
 \notag\\
 h_{BC}(\mathbf{k})
 &=t\,e^{-i\phi/3}\bigl(1+e^{+i(\mathbf{k}\cdot\mathbf{a}_1
 -\mathbf{k}\cdot\mathbf{a}_2)}\bigr),
 \label{eq:kagome_H}\\
 h_{CA}(\mathbf{k})
 &=t\,e^{-i\phi/3}\bigl(1+e^{+i\mathbf{k}\cdot\mathbf{a}_2}\bigr),
 \notag
\end{align}
with $h_{\beta\alpha}=h_{\alpha\beta}^*$ and vanishing on-site
energies.  On the kagome lattice, $d+id$-wave spinon pairing is
gauge-equivalent to $s$-wave via a sublattice-dependent gauge
rotation~\cite{Bieri2016}, so this pure-hopping ansatz with complex
phases is a valid representative of its PSG class.  All numerical
results below use $t=1$ and $\phi=0.10\pi$.  The gauge-invariant flux
per triangle and the time-reversal relation
$E(\mathbf{k},+\phi)=E(-\mathbf{k},-\phi)$ are verified explicitly.

\subsection{Band structure and Chern numbers}
\label{subsec:bands}

Diagonalization over a $300\times300$ momentum grid and evaluation of
the gauge-invariant Fukui--Hatsugai--Suzuki link-variable formula yield
band Chern numbers
\begin{equation}
 (C_1,\,C_2,\,C_3)=(-1,\;0,\;+1),
 \quad \textstyle\sum_n C_n=0,
 \label{eq:kagome_chern}
\end{equation}
with numerical values $-1.000000$, $-0.000000$, $+1.000000$.  The
occupied (lowest) band has $C_{\rm occ}=-1$ per spin species.  This
integer spinon Chern number is related to the fractional many-body
invariant $C_{\rm MB}=1/2$ measured in DMRG~\cite{Gong2014} and
VMC~\cite{Hu2015} by the parton projection $\nu_s=C/2$ derived in
Sec.~\ref{sec:parton}.

At $\phi=0$ the bare kagome nearest-neighbor model has an exactly flat
band at $E=-2t$ (band~0 in our labeling), which touches the dispersive
band~1 quadratically at the $\Gamma$ point; bands~1 and~2 are
degenerate at $K$ and $K'$, forming Dirac cones.  Turning on the flux
$\phi$ disperses the flat band and opens a Haldane-type mass at the
$K,K'$ Dirac points.  At $\phi=0.10\pi$ all three bands are dispersive,
with bandwidths $0.362$, $2.44$, $2.80$ for bands~0, 1, 2
(occupied band~0 lies at the bottom of the spectrum,
$E\in[-2.351,-1.989]\,t$).  The global gaps are $\Delta_{01}=0.362\,t$
and $\Delta_{12}=0.364\,t$, numerically close but distinct.  The
occupied band, a direct descendant of the $\phi=0$ flat band, thus
acquires its nonzero Chern number from the flux-induced dispersion
while remaining well separated from the Dirac cones formed by
bands~1--2.

\subsection{Dirac cones and band topology}
\label{subsec:cones}

At $\phi=0$, bands~1 and 2 are degenerate at $K$ and $K'$.  The
flux $\phi$ opens a Haldane-type gap; at $\phi=0.10\pi$ the Dirac mass
is $m_{\rm Dirac}=0.181\,t$.  Each valley is individually isotropic
($v_\parallel=v_\perp$ to machine precision), with $v_K=0.809\,t$ and
$v_{K'}=0.914\,t$ (between-valley ratio $0.89$).  The occupied band~0
does not participate in these Dirac cones; it is separated from band~1
by a gap of $2.80\,t$ at $K$, which persists globally
($\Delta_{01}=0.362\,t$).

The Chern number $C_{\rm occ}=-1$ is established from the full
Brillouin-zone Fukui--Hatsugai--Suzuki calculation, not from a
Dirac-cone decomposition of band~0.  The signed-cone formula
Eq.~\eqref{eq:C_signed_cones} applies to the band~1--2 Dirac cones
and correctly gives $C_2-C_1=+1$ for their Chern number difference.
The Berry curvature distribution of band~0 is shown in
Fig.~\ref{fig:kagome}(b).

\subsection{Finite-cylinder response and exponential scaling}
\label{subsec:cylinder_response}

For each YC-$L_y$ cylinder with $L_y\in\{4,6,8,10,12\}$, the
spin-transport response is computed by threading flux $\Phi$ through
the cylinder hole.  The differential polarization
$\partial P_s/\partial(\Phi/2\pi)$ is obtained from the Wilson-loop
Berry phase of the occupied band, with each sub-band phase unwrapped
continuously in $\Phi$ before differentiation.  Unwrapping is
essential: differentiating the principal branch of the Wilson-loop
phase produces numerical artifacts at branch-cut crossings.

\begin{figure*}[t]
\centering
\includegraphics[width=0.95\textwidth]{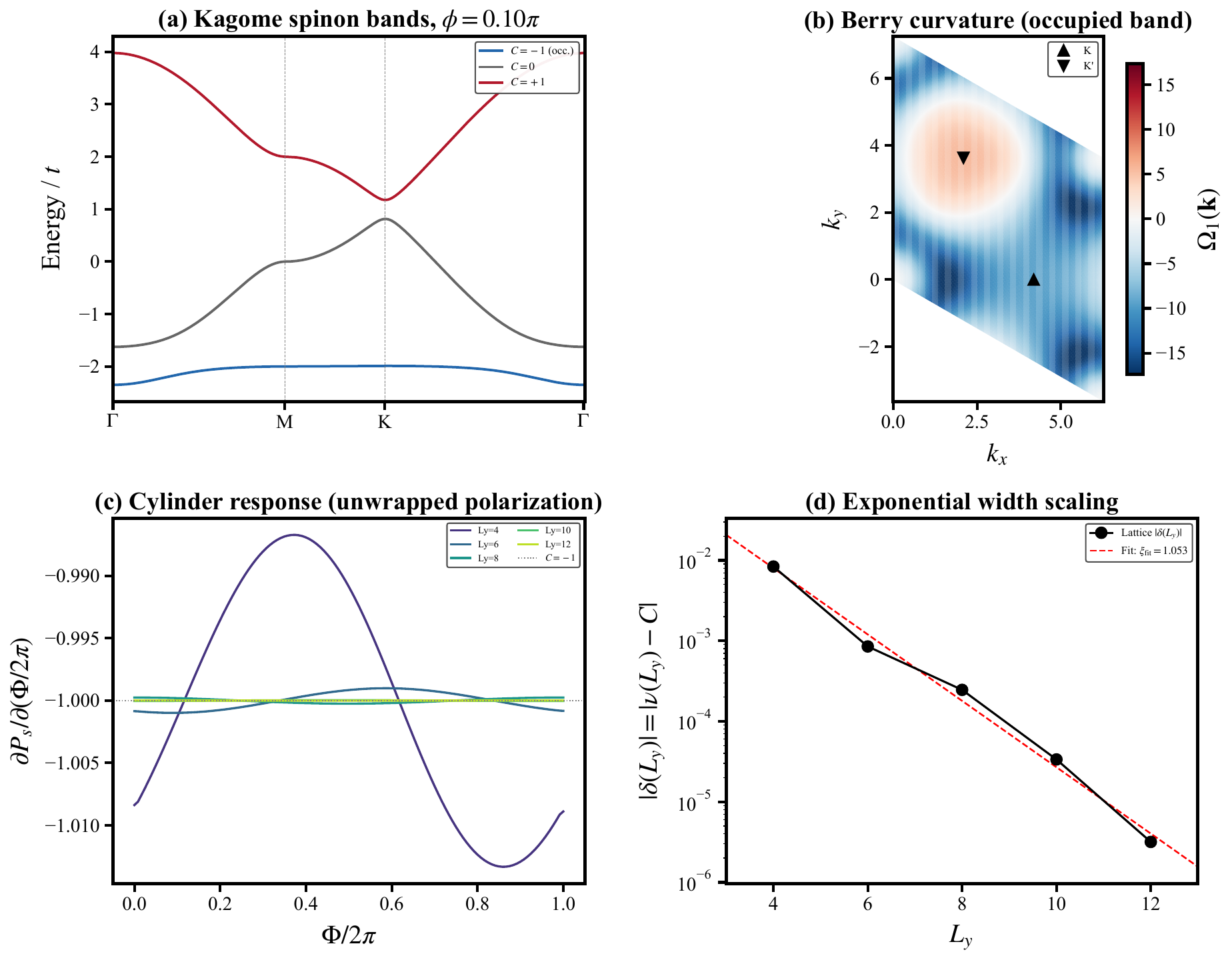}
\caption{Microscopic parton validation of the kagome chiral spin
liquid at flux $\phi=0.10\pi$ per triangle.
(a)~Spinon band structure along $\Gamma$--$M$--$K$--$\Gamma$:
occupied band (blue, $C=-1$), middle band (gray, $C=0$), upper band
(red, $C=+1$).  All three bands are dispersive at finite $\phi$.
(b)~Berry curvature $\Omega_1(\mathbf{k})$ of the occupied band, with
BZ corners $K$ (triangle) and $K'$ (inverted triangle) marked.
(c)~Differential response $\partial P_s/\partial(\Phi/2\pi)$ versus
threading flux on YC-$L_y$ cylinders for $L_y=4$--$12$, computed from
the unwrapped Wilson-loop polarization.  All curves converge to
$C_{\rm occ}=-1$.
(d)~Exponential scaling of the residual $|\delta(L_y)|$ with fitted
decay length $\xi_{\rm fit}=1.053$.  The product $\delta\cdot L_y$
(Table~\ref{tab:scaling}) is not constant, excluding universal $1/L$
corrections.}
\label{fig:kagome}
\end{figure*}

Figure~\ref{fig:kagome}(c) shows the differential response for all
five cylinder widths, converging rapidly to $C_{\rm occ}=-1$.  The
maximum deviation from $C_{\rm occ}$ is $1.3\times10^{-2}$ at $L_y=4$
and $3.3\times10^{-6}$ at $L_y=12$.  The cycle average (net winding
of the unwrapped polarization over one $2\pi$ flux period) equals
$C_{\rm occ}=-1$ to numerical precision for all $L_y$, confirming the
integer spinon Chern number.  The residual
$\delta(L_y)\equiv\partial P_s/\partial(\Phi/2\pi)\big|_{\Phi=0}
- C_{\rm occ}$
is listed in Table~\ref{tab:scaling}.

\begin{table}[b]
\caption{Finite-size residual $\delta(L_y)$ of the differential spinon
response at $\Phi=0$ on YC-$L_y$ kagome cylinders.}
\label{tab:scaling}
\begin{ruledtabular}
\begin{tabular}{rcc}
$L_y$ & $\delta(L_y)$ & $\delta\cdot L_y$ \\
\midrule
 4 & $-8.4\times10^{-3}$ & $-3.4\times10^{-2}$ \\
 6 & $-8.5\times10^{-4}$ & $-5.1\times10^{-3}$ \\
 8 & $+2.5\times10^{-4}$ & $+2.0\times10^{-3}$ \\
10 & $-3.4\times10^{-5}$ & $-3.4\times10^{-4}$ \\
12 & $+3.2\times10^{-6}$ & $+3.8\times10^{-5}$ \\
\end{tabular}
\end{ruledtabular}
\end{table}

The product $\delta\cdot L_y$ varies by three orders of magnitude and
alternates in sign, excluding a universal $1/L$ correction.  An
exponential envelope fit gives $\xi_{\rm fit}=1.053$ lattice units.
Because the occupied band does not form a Dirac cone, this effective
decay length is not directly predicted by the two-cone Poisson-kernel
formula Eq.~\eqref{eq:k_asymptotic}, which applies to bands whose
finite-size corrections are controlled by a Dirac mass.  The
exponential character of the convergence is the same, however:
Eq.~\eqref{eq:c1zero}, the analytic prediction $c_1=0$, is validated
on the microscopic lattice, confirming that the gapped bulk theory
generates no universal $1/L$ term.

\subsection{Semion CSL data}
\label{subsec:semion_data}

With $C_\uparrow=C_\downarrow=-1$, gauge projection yields the
$U(1)_{-2}$ semion theory:
\begin{equation}
 K_{\rm em}=-2,\quad
 \nu_s=-\tfrac12,\quad
 |\det K|=2,\quad
 c_-=-1.
 \label{eq:kagome_semion}
\end{equation}
The two-fold torus degeneracy agrees with the VMC overlap-matrix
construction of Hu \emph{et al.}~\cite{Hu2015}, and the modular
matrices $\mathcal S=\frac{1}{\sqrt2}
\bigl(\begin{smallmatrix}1&1\\1&-1\end{smallmatrix}\bigr)$,
$\mathcal U=e^{-2\pi i/24}
\bigl(\begin{smallmatrix}1&0\\0&i\end{smallmatrix}\bigr)$
extracted numerically by He \emph{et al.}~\cite{HeBhattacharjee2016}
match the universal data of the $K=(2C)$ theory
(Appendix~\ref{app:Kmatrix}).  He \emph{et al.} further showed that
the CSL can be understood as a gauged $U(1)$ symmetry-protected
topological phase, whose continuum Chern--Simons action reproduces
$\nu_s=C/2$ and gaps the emergent photon, ensuring
deconfinement~\cite{HeBhattacharjee2016}.  The predicted many-body
spin pump is $\Delta S^z=\nu_s=-1/2$ per $2\pi$ flux cycle, confirmed
by the pilot DMRG calculation in Sec.~\ref{sec:dmrg}.

%=======================================================================
\section{Interacting DMRG spin pump}
\label{sec:dmrg}
%=======================================================================

The parton calculation of Sec.~\ref{sec:kagome_realization} verifies
the analytic prediction $\nu_s=C/2=-1/2$ at the noninteracting spinon
level.  A complementary interacting many-body test is performed by
density-matrix renormalization group (DMRG) calculations on the
spin-$1/2$ kagome Hamiltonian
\begin{equation}
  H = J\!\sum_{\langle ij\rangle}\!\mathbf{S}_i\!\cdot\!\mathbf{S}_j
    + J_\chi\!\!\sum_{\triangle,\nabla}^{\mathrm{CCW}}\!\!
      \mathbf{S}_i\!\cdot\!(\mathbf{S}_j\!\times\!\mathbf{S}_k),
  \label{eq:spin_ham}
\end{equation}
with $J=1$ and $J_\chi/J=0.25$, well inside the CSL phase identified
by Bauer \emph{et al.}~\cite{Bauer2014} and consistent with the
independent DMRG work of Refs.~\cite{Gong2014,He2014}.

\subsection{Setup}
\label{subsec:dmrg_setup}

We use two-site DMRG with $U(1)$ $S^z$ conservation
(TeNPy~\cite{Hauschild2024}) on an $L_x=8$, $L_y=4$ cylinder ($N=96$
sites), open along $\hat{x}$ and periodic along $\hat{y}$.  A spin
flux $\Phi$ is threaded through the cylinder by a uniform twist gauge.
The bond dimension is ramped to $\chi_{\max}=600$ over the first
14 sweeps, with a density-matrix perturbation (mixer) applied at the
first flux point.  Subsequent flux points are warm-started from the
preceding converged state.  The scan covers $\Phi\in[0,4\pi]$ with
$n_\Phi=14$ equally spaced points, traversing two full Hamiltonian
periods.

\subsection{Results}
\label{subsec:dmrg_results}

\begin{figure*}[t]
\centering
\includegraphics[width=0.95\textwidth]{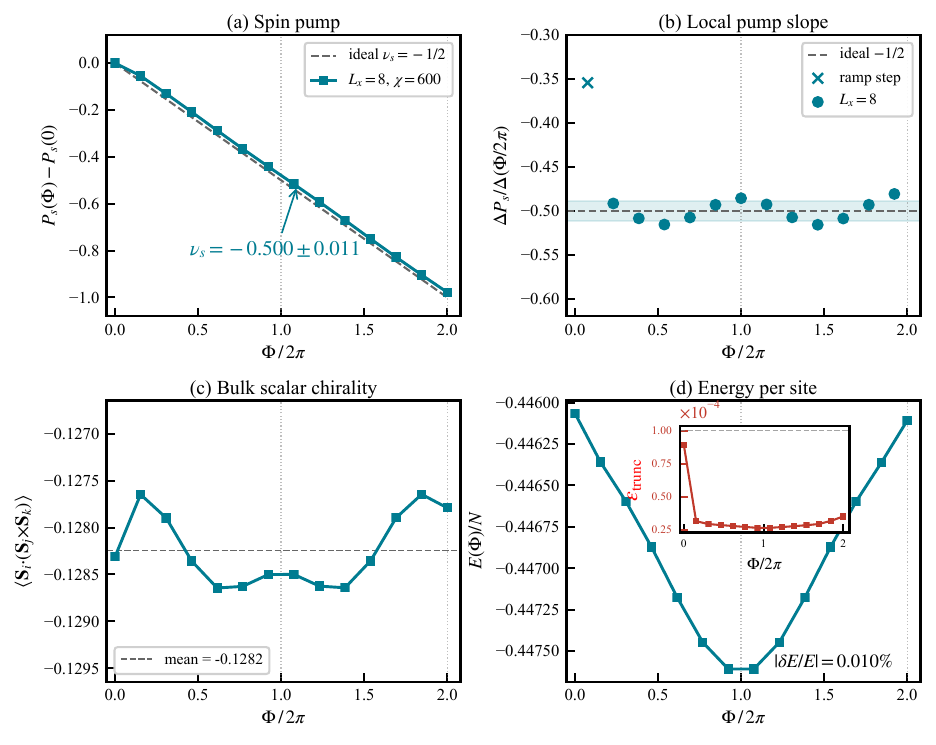}
\caption{DMRG spin pump on the kagome CSL
[Eq.~\eqref{eq:spin_ham}, $J_\chi/J=0.25$, $L_x=8$, $L_y=4$,
$\chi_{\max}=600$, $N=96$].
(a)~Accumulated pump $P_s(\Phi)-P_s(0)$; dashed: ideal $\nu_s=-1/2$.
(b)~Local slope per flux step; cross: ramp-startup step excluded from
the average; shaded band: $1\sigma$.
(c)~Bulk scalar chirality, constant to 0.3\%.
(d)~Energy per site; inset: truncation error $\epsilon_{\rm trunc}$.}
\label{fig:dmrg}
\end{figure*}

Figure~\ref{fig:dmrg} summarizes the results.
Panel~(a) shows the spin pump $P_s(\Phi)-P_s(0)$ over two flux
periods.  The curve is linear throughout, reaching
\begin{equation}
  P_s(2\pi)=-0.516,
  \qquad
  P_s(4\pi)=-0.978,
  \label{eq:dmrg_pump}
\end{equation}
bracketing the ideal quantized values $-1/2$ and $-1$.  Panel~(b)
displays the local pump slope per flux step.  Excluding the first
step, where the bond dimension has not yet reached $\chi_{\max}$, the
12 remaining slopes give
\begin{equation}
  \nu_s=-0.500\pm0.011,
  \label{eq:nus_dmrg}
\end{equation}
in agreement with the Chern--Simons prediction $\nu_s=C/2=-1/2$.
Panel~(c) shows the bulk scalar chirality
$\kappa=\langle\mathbf{S}_i\cdot(\mathbf{S}_j\times\mathbf{S}_k)\rangle
=-0.1282\pm0.0004$ (0.3\% relative variation across the scan),
confirming that the system remains in the CSL phase without a level
crossing.  Panel~(d) shows the energy per site with a truncation-error
inset: the energy returns to its initial value after two periods with
$|\delta E/E|=9.8\times10^{-3}\%$, and time-reversal symmetry
$E(\Phi)=E(4\pi-\Phi)$ is satisfied to sub-ppm accuracy across all
paired flux points; truncation errors remain in the range
$2.7$--$8.9\times10^{-5}$ throughout.

\subsection{Interpretation}
\label{subsec:dmrg_interp}

Equation~\eqref{eq:nus_dmrg} confirms, without reference to the
parton construction, that Eq.~\eqref{eq:spin_ham} at $J_\chi/J=0.25$
realizes a ground state with $\nu_s=-1/2$.  The $4\pi$ periodicity
$P_s(4\pi)\approx2\,P_s(2\pi)$ is the semionic $\mathbb Z_2$
periodicity: the many-body ground state returns to itself only after
two flux quanta.  Together with the noninteracting parton calculation
of Sec.~\ref{sec:kagome_realization} and the field-theory derivation
of Secs.~\ref{sec:direct}--\ref{sec:parton},
Eq.~\eqref{eq:nus_dmrg} constitutes a three-way confirmation that the
topological content of the kagome CSL is the $U(1)_{-2}$ Chern--Simons
theory with $\nu_s=-1/2$, robust to interactions beyond the mean-field
saddle point.  These results use a single cylinder width $L_y=4$; a
width-resolved scan at $L_y=4,6,8$ with bond-dimension extrapolation
would provide a fully controlled finite-size analysis, and the
protocol appropriate to such a scan is described in
Appendix~\ref{app:finite_size}.

%=======================================================================
\section{Discussion and outlook}
\label{sec:discussion}
%=======================================================================

The parity anomaly of a $(2+1)$-dimensional Dirac cone does not by
itself determine the topological order of a magnet.  It determines
the half-integer contribution of each low-energy massive cone and the
integer jump under a mass inversion.  The full spinon Chern number
must be computed from a gauge-invariant lattice Hamiltonian or
Green function; the parton projection then determines the
topological field theory; and the many-body Chern number and flux
pump diagnose the interacting spin model.  This paper assembles these
four layers into a single quantitative chain and closes it on the
kagome CSL.

Equation~\eqref{eq:K_full} is the complete mass-dependent one-loop
parity-odd two-point kernel for arbitrary Euclidean momentum and
compact harmonic; Eq.~\eqref{eq:Pi_regulated} adds the only allowed
local ultraviolet term.  For a quantized transverse-flux background,
Eq.~\eqref{eq:Gamma_exact_full} is exact to all orders in the compact
holonomy and has the correct large-gauge winding.  These statements do
not constitute a closed expression for the fully nonlinear determinant
in an arbitrary space- and time-dependent field, and no such stronger
claim is required for the static cylinder response considered here.
The factor-of-two prescription of Eq.~\eqref{eq:CS_restricted}, the
explicit anomaly branch of Eq.~\eqref{eq:Gamma_exact_full}, and the
Green-function bridge of Eq.~\eqref{eq:Green_invariant} give the exact
holonomy derivative and the local polarization tensor identical
normalization, prevent the mass-dependent arctangent from being
mistaken for a separately gauge-invariant action, and connect the
Coste--Luescher regulator integer to the remote-band contribution to
the physical Chern number.

The kagome parton calculation of Sec.~\ref{sec:kagome_realization}
establishes the Chern number $C_{\rm occ}=-1$ from a
Brillouin-zone lattice invariant and verifies exponential convergence
of the finite-cylinder response with $c_1=0$.  The Dirac cones of the
three-band model sit between bands~1 and~2 (with mass
$m_{\rm Dirac}=0.181\,t$ at $\phi=0.10\pi$), while the occupied band~0
is separated by a larger gap; the two-cone Poisson-kernel formula
Eq.~\eqref{eq:k_asymptotic} therefore does not directly predict the
finite-size corrections of band~0.  The exponential character of the
convergence and the absence of any $1/L$ term nevertheless follow from
the general finite-size argument of Sec.~\ref{sec:direct}, since the
gapped bulk generates only exponential corrections regardless of the
microscopic origin of the gap.  A parallel taxonomy of three
inequivalent finite-size scalings, and the distinction between the
bulk pump and the algebraic $1/\ell$ scaling of a chiral edge, is
collected in Appendix~\ref{app:finite_size}.

The pilot DMRG calculation of Sec.~\ref{sec:dmrg} closes the same
chain at the interacting many-body level: the spin pump on the
explicitly chiral kagome model gives $\nu_s=-0.500\pm0.011$, in precise
agreement with the Chern--Simons prediction.  The $4\pi$ periodicity
and the uniform scalar chirality throughout the flux scan show that
the measurement is performed entirely within the CSL phase.  Combining
the three levels of description, the field-theory
derivation~(Secs.~\ref{sec:direct}--\ref{sec:parton}), the
noninteracting parton band
calculation~(Sec.~\ref{sec:kagome_realization}), and the interacting
DMRG pump~(Sec.~\ref{sec:dmrg}), yields a quantitative bridge from
microscopic topology to observable fractional response with no
adjustable parameters.

Several extensions are natural.  A width-resolved DMRG scan at
$L_y=4,6,8$ with explicit bond-dimension extrapolation, following the
protocol of Appendix~\ref{app:finite_size}, would allow a direct
comparison between the MPS correlation length $\xi_{\rm MPS}$ and the
parton decay length $\xi_{\rm fit}=1.053$.  The same three-level
strategy applies without modification to other lattice CSL candidates
whose parton bands are Chern insulators, and to the identification of
$U(1)_{2C}$ semion theories with $|C|\ge2$ that would arise from
Chern bands with higher $|C|$.  The distinction between the semionic
$U(1)_2$ response studied here and the Ising response of the
non-Abelian Kitaev phase (Appendix~\ref{app:kitaev}) provides a sharp
diagnostic for distinguishing candidate spin liquids in experiment,
because the semion CSL has a spin $U(1)$ pump equal to $C/2$ while the
Kitaev phase has only a gravitational (thermal) analogue.
The four-dimensional CPT anomaly of Ghosh and
Klinkhamer~\cite{GhoshKlinkhamer2018} is not the relevant ultraviolet
completion here: the same regulator construction evaluated with the
native two-dimensional loop measure of the present problem yields
$1/(ML)$ scaling that vanishes as $M\to\infty$, so the ultraviolet
integer arises from the three-dimensional parity anomaly and the
Bloch/Green-function Chern invariant, not from a four-dimensional
mechanism.

%=======================================================================
\bibliographystyle{apsrev4-2}
\bibliography{ref}
%=======================================================================

%=======================================================================
\appendix
\onecolumngrid
%=======================================================================

\section{Full parity-odd kernel from Feynman parametrization}
\label{app:full_kernel}

For canonical velocities and the minimally normalized field, the
quadratic loop is
\begin{equation}
 \Pi^{\mu\nu}(p)=-\frac{1}{L_{\rm eff}}\sum_n
 \int\frac{\dd^2k}{(2\pi)^2}
 \tr\!\left[
 \gamma^\mu\frac{-i\slashed{k}+m}{k^2+m^2}
 \gamma^\nu\frac{-i(\slashed{k}+\slashed{p})+m}
 {(k+p)^2+m^2}\right].
 \label{eq:app_Pi}
\end{equation}
With the trace orientation chosen to agree with the occupied-band
convention in Appendix~\ref{app:DiracChern}, the odd numerator is
\begin{equation}
 N_{\rm odd}^{\mu\nu}=2i\chi m\epsilon^{\mu\nu\rho}p_\rho.
 \label{eq:app_Nodd}
\end{equation}
The loop-momentum pieces cancel before integration.  Feynman
parametrization and integration over the two noncompact loop components
give
\begin{equation}
 \Pi_{\rm odd}^{\mu\nu}(p)=
 \frac{i\chi m}{2\pi L_{\rm eff}}\epsilon^{\mu\nu\rho}p_\rho
 \int_0^1\dd u\sum_{n\in\mathbb Z}
 \frac{1}{(\kappa_n+up_y)^2+\Delta_u^2},
 \label{eq:app_kernel_sum}
\end{equation}
where
\begin{equation}
 \kappa_n=\frac{2\pi n+\vartheta}{L_{\rm eff}},\qquad
 p_y=\frac{2\pi r}{L_{\rm eff}},\qquad
 \Delta_u^2=m^2+u(1-u)p^2.
 \label{eq:app_kappa_Delta}
\end{equation}
The exact sum
\begin{equation}
 \frac{1}{L_{\rm eff}}\sum_n
 \frac{1}{(\kappa_n+up_y)^2+\Delta_u^2}
 =\frac{1}{2\Delta_u}
 \frac{\sinh(L_{\rm eff}\Delta_u)}
 {\cosh(L_{\rm eff}\Delta_u)-\cos(\vartheta+2\pi r u)}
 \label{eq:app_shifted_sum}
\end{equation}
yields Eq.~\eqref{eq:K_full}.  Restoring the original probe field
multiplies the answer by $q^2$.

For anisotropic velocities, set
$K_x=|v_\perp|k_x$ and $K_y=|v_\parallel|k_y$.  The loop measure
contributes $1/(|v_\perp v_\parallel|)$, while the two current
vertices and the external momentum in the odd trace contribute
$|v_\perp v_\parallel|$; their magnitudes cancel.  The remaining
orientation sign, including the gamma-matrix convention, is precisely
$\chi$, and the compact spacing becomes $2\pi/L_{\rm eff}$ with
$L_{\rm eff}=L/|v_\parallel|$.  This proves the anisotropic form used
in the main text.  A gauge-invariant regularization can change the
answer only by a local transverse odd tensor.  Adding
$n+\chi\eta/2$ to the scalar form factor gives
Eq.~\eqref{eq:Pi_regulated}; it does not alter any nonlocal momentum
dependence.

\section{Restricted Chern--Simons normalization and holonomy primitive}
\label{app:holonomy}

For a general field,
\begin{equation}
 \int b\wedge\dd b
 =\int\dd^3x\,\epsilon^{\mu\nu\rho}b_\mu\partial_\nu b_\rho.
 \label{eq:app_forms}
\end{equation}
To evaluate this functional on constant $b_y$, one must first retain
a small transverse momentum in $b_y$ and integrate by parts.  The
terms $b_yF_{\tau x}$ and
$b_\tau\partial_xb_y-b_x\partial_\tau b_y$ then give equal limits.
Thus
\begin{equation}
 \frac{i k}{4\pi}\int b\wedge\dd b
 \longrightarrow i k
 \left(\frac{1}{2\pi}\int_{\Sigma_2}\dd b\right)
 \left(\oint_{S^1}b\right)=ik\mathcal N\Theta.
 \label{eq:app_CS_factor_two}
\end{equation}
Substituting a strictly constant $b_y$ before this limiting operation
would miss the second contribution and give an incorrect factor of two.

The determinant can also be obtained directly from the compact-mode
factorization.  The $n$th mode is a two-dimensional Dirac operator
with complex mass $m+i\gamma^y\kappa_n$.  A constant chiral rotation
removes its phase and produces the two-dimensional anomalous Jacobian.
Relative to zero holonomy, the mass-odd phase is therefore the
symmetrically regulated sum
\begin{align}
 \Gamma_{\rm odd}^{\rm IR}(\Theta)-\Gamma_{\rm odd}^{\rm IR}(0)
 =i\chi\mathcal N\sum_{n\in\mathbb Z}\Bigg[&
 \arctan\frac{2\pi n+\alpha+\Theta}{mL_{\rm eff}}\notag\\
 &-\arctan\frac{2\pi n+\alpha}{mL_{\rm eff}}\Bigg].
 \label{eq:app_mode_phase_sum}
\end{align}
The difference is defined unambiguously by differentiating first:
\begin{align}
 \frac{1}{i\chi\mathcal N}\frac{\partial\Gamma_{\rm odd}^{\rm IR}}
 {\partial\Theta}
 &=\frac{m}{L_{\rm eff}}\sum_n
 \frac{1}{m^2+[(2\pi n+\alpha+\Theta)/L_{\rm eff}]^2}\notag\\
 &=\frac12\frac{\sinh(mL_{\rm eff})}
 {\cosh(mL_{\rm eff})-\cos(\alpha+\Theta)}.
 \label{eq:app_mode_sum_derivative}
\end{align}
Integrating from $0$ to $\Theta$ fixes the normalization and gives
Eq.~\eqref{eq:Gamma_exact_IR}.

For the two most common spin structures the primitive
Eq.~\eqref{eq:A_holonomy_closed} reduces to
\begin{align}
 \mathcal A_{\pi}(x,\Theta)
 &=\operatorname{Arg}_{\rm cont}
 \left[\cos\frac{\Theta}{2}
 +i\tanh\frac{x}{2}\sin\frac{\Theta}{2}\right],
 \label{eq:A_antiperiodic}\\
 \mathcal A_{0}(x,\Theta)
 &=\operatorname{Arg}_{\rm cont}
 \left[\cos\frac{\Theta}{2}
 +i\coth\frac{x}{2}\sin\frac{\Theta}{2}\right].
 \label{eq:A_periodic}
\end{align}
The antiperiodic derivative at zero holonomy reduces to the familiar
$\tanh(|m|L/2)$ finite-compactification
coefficient~\cite{BabuDasPanigrahi1987}.  The periodic expression is
singular at $m=0$, as required by the compact zero mode.

The derivative identity
\begin{equation}
 \frac{\dd}{\dd\Theta}
 \left[\arctan\!\left(
 \coth\frac{x}{2}\tan\frac{\alpha+\Theta}{2}\right)
 \right]_{\rm cont}
 =\frac12\frac{\sinh x}{\cosh x-\cos(\alpha+\Theta)}
 \label{eq:app_primitive}
\end{equation}
proves Eqs.~\eqref{eq:A_holonomy_integral} and
\eqref{eq:A_holonomy_closed}.  Integrating over one cycle gives
\begin{equation}
 \int_0^{2\pi}\frac{\dd\Theta}{2}
 \frac{\sinh x}{\cosh x-\cos(\alpha+\Theta)}
 =\pi\sgn(x),
 \label{eq:app_winding_integral}
\end{equation}
yielding Eq.~\eqref{eq:A_winding}.  Finally,
\begin{equation}
 \frac{\sinh\lambda}{\cosh\lambda-\cos\vartheta}
 =1+2\sum_{r=1}^\infty e^{-r\lambda}\cos(r\vartheta)
 \label{eq:Poisson_kernel}
\end{equation}
is the Poisson kernel: it proves both the exact cycle average and the
exponential winding expansion.  Each harmonic represents a virtual
massive trajectory winding around the cylinder.

\section{Green-function reduction and Wilson-degree formula}
\label{app:GreenChern}

For $G^{-1}(i\omega,\bm k)=i\omega-h(\bm k)$ with a spectral gap,
insert the spectral resolution
$h=\sum_nE_n|u_n\rangle\langle u_n|$ into
Eq.~\eqref{eq:Green_invariant}.  The frequency integral can be closed
in the upper or lower half-plane; only poles separated by the Fermi
level contribute, and the result reduces to
\begin{equation}
 N_3[G]=\frac{1}{2\pi}\sum_{n\in{\rm occ}}
 \int_{\rm BZ}\dd^2k\,
 i\epsilon^{ij}\langle\partial_i u_n|\partial_j u_n\rangle
 =C,
 \label{eq:app_Green_to_Chern}
\end{equation}
which is Eq.~\eqref{eq:lattice_Chern}.  This is the precise bridge
between the three-dimensional propagator winding and the two-dimensional
occupied-band invariant.

In the three-dimensional Euclidean Wilson regulator, let $Q(p)$ be the
nonsingular two-component kernel on the Brillouin torus.  Its
normalized map
\begin{equation}
 U(p)=\frac{Q(p)}{\sqrt{\det Q(p)}}\in SU(2)
 \label{eq:normalized_wilson_map}
\end{equation}
has degree
\begin{equation}
 n[U]=\frac{1}{24\pi^2}\int_{T^3_{\rm BZ}}
 \tr[(U^{-1}\dd U)^3]\in\mathbb Z.
 \label{eq:propagator_winding}
\end{equation}
Coste and Luescher~\cite{CosteLuscher1989} found admissible Wilson
signs in the $n=0$ and $n=-1$ classes despite identical naive
low-energy propagators, showing that a formally irrelevant ultraviolet
term can select a distinct parity-anomaly class.  A physical flat band
cannot by itself be identified with a Wilson doubler or with a
Ginsparg--Wilson construction; such a claim would require an explicit
local overlap operator, a chirality operator, a spectral gap, and a
locality proof.  The Bloch/Green-function Chern invariant is the
appropriate nonperturbative completion for the CSL problem.

\section{Signed Dirac-cone Chern number}
\label{app:DiracChern}

For the two-band Bloch Hamiltonian
\begin{equation}
 h_v(\bm q)=v_{x,v}q_x\sigma_x+v_{y,v}q_y\sigma_y+m_{H,v}\sigma_z,
 \label{eq:app_hDirac}
\end{equation}
the occupied-band Berry curvature is
\begin{equation}
 \Omega_v(\bm q)=
 -\frac{m_{H,v}v_{x,v}v_{y,v}}
 {2(v_{x,v}^2q_x^2+v_{y,v}^2q_y^2+m_{H,v}^2)^{3/2}}.
 \label{eq:app_BerryDirac}
\end{equation}
Therefore
\begin{equation}
 \frac{1}{2\pi}\int\dd^2q\,\Omega_v
 =-\frac12\sgn(v_{x,v}v_{y,v}m_{H,v}).
 \label{eq:app_halfChern_raw}
\end{equation}
The Euclidean mass parameter and trace orientation are fixed so that
\begin{equation}
 \chi_v\sgn(m_v)
 \equiv-\sgn(v_{x,v}v_{y,v}m_{H,v}),
 \label{eq:app_chi_mapping}
\end{equation}
and hence
\begin{equation}
 C_v=\frac12\chi_v\sgn(m_v).
 \label{eq:app_halfChern}
\end{equation}
This explicit mapping prevents a sign convention in the Euclidean
gamma matrices from being mistaken for a physical change of
chirality.  A full lattice calculation is still needed to determine
the ultraviolet integer and to exclude additional gap closings.

\section{Semion $K$ matrix and universal data}
\label{app:Kmatrix}

Starting from Eq.~\eqref{eq:parton_hydrodynamic}, the constraint sets
$\alpha_\downarrow=-\alpha_\uparrow\equiv-b$ and gives
\begin{equation}
 S_{\rm top}=\frac{i}{4\pi}\int 2C\,b\dd b
 +\frac{i}{2\pi}\int A^s\dd b.
 \label{eq:app_K_action}
\end{equation}
Thus $K=(2C)$ and $t=(1)$.  For quasiparticle label $\ell$,
\begin{align}
 Q_s(\ell)&=t^TK^{-1}\ell=\frac{\ell}{2C}\quad(\mathrm{mod}\ 1),
 \label{eq:app_spin_charge}\\
 \theta_\ell&=\pi\ell^TK^{-1}\ell=\frac{\pi\ell^2}{2C},
 \label{eq:app_statistics}\\
 \theta_{\ell\ell'}&=2\pi\ell^TK^{-1}\ell'
 =\frac{\pi\ell\ell'}{C}.
 \label{eq:app_mutual}
\end{align}
For $C=1$, the nontrivial quasiparticle $\ell=1$ has spin $1/2$ modulo
an integer and exchange phase $e^{i\pi/2}$.  The genus-$g$ degeneracy
is
\begin{equation}
 \mathcal D_g=|\det K|^g=2^g,
 \label{eq:app_degeneracy}
\end{equation}
and the response invariant and central charge are
\begin{equation}
 \nu_s=t^TK^{-1}t=\frac{1}{2C},\qquad
 c_-=\operatorname{signature}(K)=\sgn C.
 \label{eq:app_nu}
\end{equation}

\section{Discrete many-body Chern number for numerical work}
\label{app:discreteChern}

For a mesh of twists $\bm\theta_{mn}$, a gauge-invariant
discretization uses overlap matrices between neighboring ground-state
multiplets,
\begin{equation}
 [M_i(\bm\theta)]_{ab}=\langle\Psi_a(\bm\theta)|
 \Psi_b(\bm\theta+\Delta\theta_i\hat i)\rangle.
 \label{eq:app_overlap}
\end{equation}
Define link variables
\begin{equation}
 U_i(\bm\theta)=\frac{\det M_i(\bm\theta)}{|\det M_i(\bm\theta)|}
 \label{eq:app_link}
\end{equation}
and plaquette curvature
\begin{equation}
 F_{xy}(\bm\theta)=\operatorname{Arg}
 \left[U_x(\bm\theta)U_y(\bm\theta+\Delta\theta_x)
 U_x^{-1}(\bm\theta+\Delta\theta_y)U_y^{-1}(\bm\theta)\right].
 \label{eq:app_plaquette}
\end{equation}
Then
\begin{equation}
 C_{\rm MB}=\frac{1}{2\pi}\sum_{mn}F_{xy}(\bm\theta_{mn})
 \label{eq:app_discrete_C}
\end{equation}
is integer when the ground-state multiplet remains separated from
excited states over the twist torus.  This supplies a direct
exact-diagonalization or finite-DMRG cross-check of the cylinder
pump.  For a finite twist mesh, the discrete analogue of the
fixed-$\theta_x$ pump is obtained by summing plaquette phases along
one strip; it can vary with the strip at finite size, but summing all
strips gives $2\pi C_{\rm MB}$ exactly (provided no overlap
determinant vanishes), the lattice version of
Eq.~\eqref{eq:pump_twist_average}.

\section{Kagome benchmarks and Hubbard origin}
\label{app:benchmarks}

Reference~\cite{Gong2014} on the $J$--$J'$ model
[Eq.~\eqref{eq:gong_model}] displayed a quantized pump primarily for
$L_y=4$, with $L_y=4$ and $6$ compared for correlation functions and
topological sector diagnostics rather than as an independent
Hall-response scan; convergence near phase boundaries required
substantially larger bond dimensions than deep in the CSL.  A
width-resolved calculation at a point such as $J'/J=0.5$ well inside
the phase, with explicit bond-dimension extrapolation, is the
appropriate next step.  Reference~\cite{Bauer2014} on the explicitly
chiral model [Eq.~\eqref{eq:bauer_model}] gives the conservative
thermodynamic bound
\begin{equation}
 0<\left(\frac{J_\chi}{J_{\rm HB}}\right)_{\rm crit}<0.16,
 \label{eq:bauer_critical_bound}
\end{equation}
and shows stability against Dzyaloshinskii--Moriya coupling of order
the bulk gap, next-nearest-neighbour exchange in the tested interval
$J_{\rm NNN}\in[-0.1,0.27]$ at $\theta=0.15\pi$, and a substantial
range of easy-axis anisotropy.  A controlled DMRG test therefore uses
a point safely inside the CSL, for example
$J_\chi/J_{\rm HB}\simeq0.2$--$0.3$, and verifies that the measured
bulk gap and correlation length remain stable as the width grows,
matching the choice $J_\chi/J=0.25$ used in
Sec.~\ref{sec:dmrg}.

The supplement of Ref.~\cite{Bauer2014} also connects
Eq.~\eqref{eq:bauer_model} to a half-filled Hubbard model with
dimensionless Peierls phase $\Phi$ through each triangle.  For
$t/U\ll1$,
\begin{align}
 J_{\rm HB}
 &=\frac{4t^2}{U}\left(1-\frac{32t^2}{U^2}\right)+\cdots,
 \label{eq:JHB_hubbard}\\
 J_\chi&=\frac{24\Phi t^3}{U^2}+\cdots,
 \label{eq:Jchi_hubbard}
\end{align}
so that
\begin{equation}
 \frac{J_\chi}{J_{\rm HB}}
 =6\Phi\frac{t}{U}
 \left(1-\frac{32t^2}{U^2}\right)^{-1}+\cdots.
 \label{eq:ratio_hubbard}
\end{equation}
This is a microscopic bridge between orbital flux, the Haldane-type
spinon mass, and the induced Chern--Simons response.
Equations~\eqref{eq:JHB_hubbard}--\eqref{eq:ratio_hubbard} are
controlled only in the strong-coupling regime; quantitative material
estimates near the Mott transition require simulations of the full
Hubbard model rather than an extrapolation of the truncated series.

\section{Kitaev phase as distinct topological response}
\label{app:kitaev}

The field-induced non-Abelian Kitaev phase is not a $U(1)_1$ analogue
of the semion CSL~\cite{Kitaev2006,ReadGreen2000}.  Its gapped
Majorana band has integer Chern number $\nu_M=\pm1$, producing Ising
topological order and a chiral Majorana edge with
\begin{equation}
 c_-=\frac{\nu_M}{2}=\pm\frac12.
 \label{eq:Kitaev_c}
\end{equation}
Because generic Kitaev interactions do not conserve $S^z$, the spin
$U(1)$ flux pump defined in Sec.~\ref{sec:manybody} is not generally
available.  The robust response is gravitational/thermal,
\begin{equation}
 \frac{\kappa_{xy}}{T}=c_-\frac{\pi^2k_B^2}{3h}.
 \label{eq:thermal_general}
\end{equation}
This response is distinct from the $U(1)_2$ semion signature analyzed
in the main text and constitutes an independent diagnostic in
experiment.

\section{Three finite-size scalings and the DMRG fitting protocol}
\label{app:finite_size}

Three finite-size phenomena in a topologically ordered phase obey
different scaling laws and must be analyzed separately.

\emph{Bulk topological-sector splitting.}  On an infinite cylinder,
minimally entangled states carrying different anyon flux are distinct
superselection sectors; differences of local bulk observables and
energy densities vanish exponentially with circumference.  On a finite
torus, anyon tunneling around either noncontractible cycle produces
\begin{align}
 \Delta E_{\rm topo}
 ={}&A_x e^{-L_x/\xi_x}\cos(k_xL_x+\varphi_x)\notag\\
 &+A_y e^{-L_y/\xi_y}\cos(k_yL_y+\varphi_y)+\cdots.
 \label{eq:topo_split}
\end{align}
On a finite open cylinder, boundary conditions can additionally select
or mix sectors, with end-to-end effects exponentially small in the
cylinder length when the edges are gapped or well separated.  None of
these bulk topological splittings has a universal $1/L$ form.

\emph{Bulk Hall response and flux pumping.}  The free massive-cone
calculation predicts Eq.~\eqref{eq:k_asymptotic} for the differential
spinon response, verified on the lattice in
Sec.~\ref{sec:kagome_realization}.  For an interacting gapped phase,
quasi-adiabatic continuation and finite correlation length make
exponential convergence the natural null model for a fixed-twist
pump or differential polarization response $X(L)$:
\begin{equation}
 X(L)=X(\infty)+B e^{-L/\xi_{\rm eff}}+\cdots.
 \label{eq:nu_exp}
\end{equation}
The coefficient and effective length are nonuniversal and can depend
on cylinder orientation, topological sector, and twist protocol.  The
one-loop result predicts a twist-sensitive sign through
$\cos\vartheta_v$, but gauge projection and interactions can modify
the amplitude and combine several correlation lengths.  A
phenomenological $1/L$ fit may still be compared with data, but it is
not the asymptotic prediction of the gapped bulk theory.

\emph{Chiral edge level spacing.}  A genuine algebraic $1/\ell$ law
occurs for a chiral edge of finite perimeter $\ell$.  Conformal
invariance gives
\begin{align}
 E_n&=e_\infty\ell+
 \frac{2\pi v_{\rm edge}}{\ell}
 \left(h_n-\frac{c_-}{24}\right)+\order{\ell^{-2}},
 \label{eq:edge_CFT}\\
 E_n-E_0&=\frac{2\pi v_{\rm edge}}{\ell}h_n+\cdots.
 \label{eq:edge_gap}
\end{align}
For the semion CSL, the edge is the chiral $SU(2)_1$ or equivalently
compact-boson theory with $c_-=1$.  Bauer \emph{et al.} observed this
$c=1$ edge spectrum while finding a finite bulk gap on
cylinders~\cite{Bauer2014}.  Equation~\eqref{eq:edge_CFT}, rather than
topological-sector splitting, is the appropriate source of universal
$1/\ell$ scaling.

\emph{Fitting protocol.}  For a width-resolved fixed-twist pump or
differential response $X(L)$, the statistically neutral comparison
is
\begin{align}
 \mathcal M_{\rm exp}:\quad
 X(L)&=X_\infty+B e^{-L/\xi},
 \label{eq:model_exp}\\
 \mathcal M_{1/L}:\quad
 X(L)&=X_\infty+\frac{A}{L}.
 \label{eq:model_invL}
\end{align}
The fit should include digitization uncertainty, DMRG truncation
error, and covariance among points obtained from the same flux sweep.
Model comparison should report corrected Akaike information criterion
or leave-one-width-out prediction error, rather than selecting the
visually best curve.  For the integrated physical spin pump, the
thermodynamic value $X_\infty=\pm1/2$ should be tested both as fixed
and free.  A local spinon slope has a different normalization and
must not be fitted to the fractional many-body value without the
parton projection.  A width-resolved DMRG scan should keep $L_x$ large
enough that edge profiles are separated, compare $L_y=4,6,8$ (and
$10$ if feasible), and repeat each width at several bond dimensions;
the full-cycle pump and the local slope
$\partial P_s/\partial(\Phi/2\pi)$ should be reported separately.  The
pilot calculation in Sec.~\ref{sec:dmrg} implements this protocol at
a single width $L_y=4$.

\end{document}